\documentclass[prb,notitlepage]{revtex4-1} 

\usepackage{amsmath} 
\usepackage{amsfonts}
\usepackage{graphicx}
\usepackage{color}
\usepackage{float}
\usepackage{leftindex}
\usepackage{gensymb}
\usepackage[dvipsnames]{xcolor}

\begin{document}

\title{Attraction of a jerk}

\date{April 1st 2025}

\author{Sini Peltonen} 
\author{Laura Vasko}
\author{Inka Tenhunen}
\author{Inka Hiltunen}
\author{Ronja Laurén}
\author{Jami J. Kinnunen}
\email{jami.kinnunen@aalto.fi}
\affiliation{Department of Applied Physics, Aalto University School of Science, FI-00076 Aalto, Finland}

\begin{abstract}
Jerk plays a pivotal role in the thrilling experience of many amusemement park rides. In addition to exploring the physical aspect of jerks, we tackle the empirical observation of an attractive force between passengers in the popular attraction, the spinning teacups. By modeling the complex system of rotating platforms, we show that pseudotorques induced by changing acceleration lead to jerky movements and an attractive interaction among riders. Our numerical analysis confirms the empirical observations, highlighting the connection between attraction and jerks.
\end{abstract}

\maketitle 

\section{Introduction} 

In this manuscript we study the popular attraction found in nearly all amusement parks, the spinning teacups. More precisely, we explore the different jerks one encounters on this ride. Not the jerks that cut you in the line, but the jerks that make you scream of happiness. After all that is what makes amusement parks attractive to the visitors: acceleration and the changes in acceleration, jerks. 

Unlike social sciences, the physics textbooks seem to have a noticeable lack of information on jerks~\cite{Young2020}. Overall, jerk is a subject that offers opportunities for further study. In the years 2015 and 2020 a total number of 202 articles on jerk in the context of linear dynamics were published~\cite{pendrill-14-9-2020}. 
Interestingly, there's no clear consensus on how much jerkiness the human body can tolerate; it's a trial-and-error situation for thrill-seekers to determine when to leave the jerk behind.~\cite{pendrill2020}

Empirical observations suggest a peculiar phenomenon during the teacup rides: an attractive force between the riders within the same teacup, so-called teamates. The pressing question is whether this force is a result of the carousel's motion, or is it simply the magnetic personalities of the occupants? In this article, we choose to model the former.

\begin{figure}[h!]
    \centering \includegraphics[width=0.4\columnwidth]{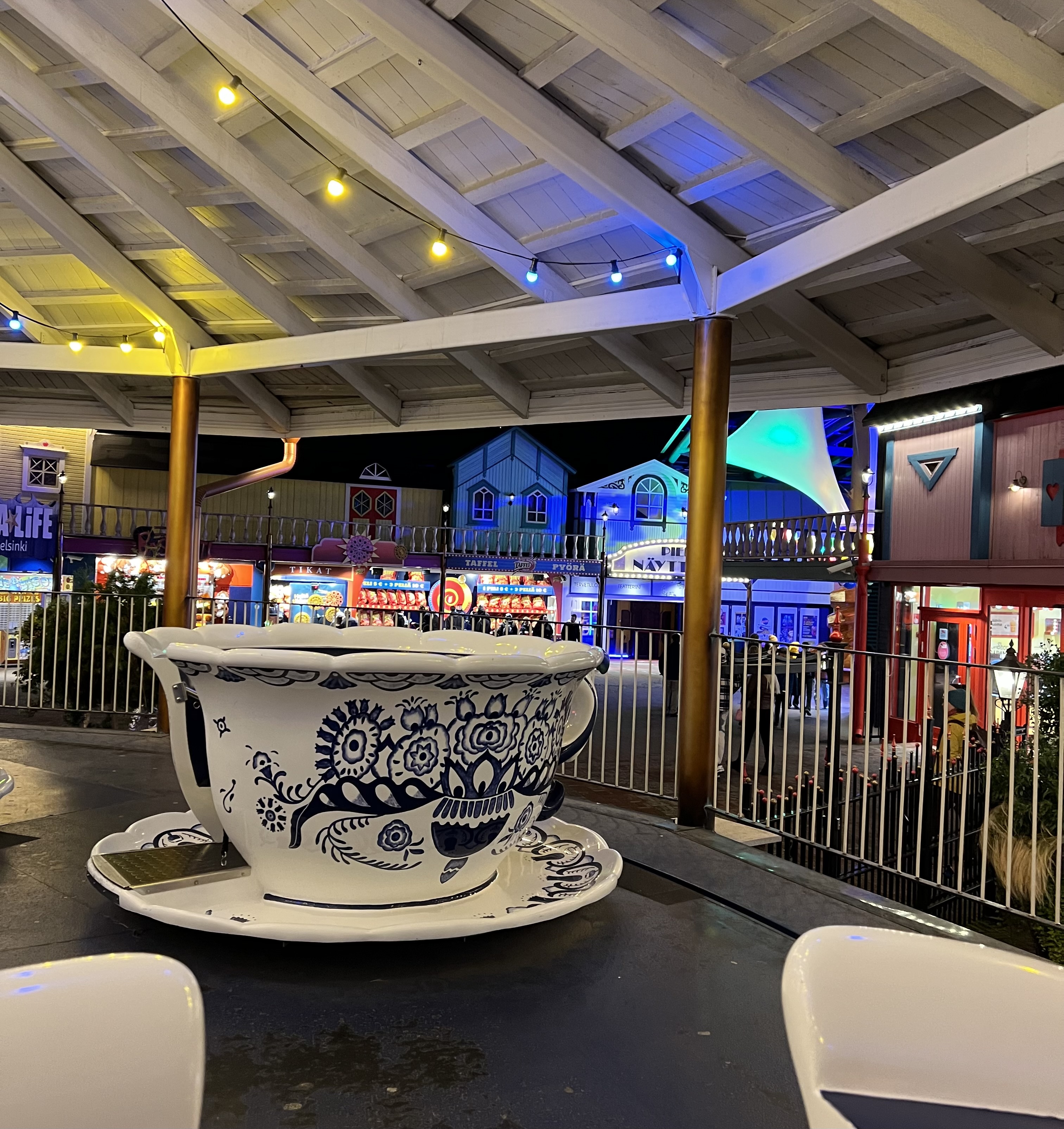}
    \caption{One can feel the attraction when sitting in a cup. (Inka Tenhunen, 11.10.2024, Linnanmäki, Helsinki)}
    \label{fig:arago_formation}
\end{figure}

\section{Teacup-teamate complex}

\begin{figure}[H]
    \centering
    \includegraphics[width=0.5\linewidth]{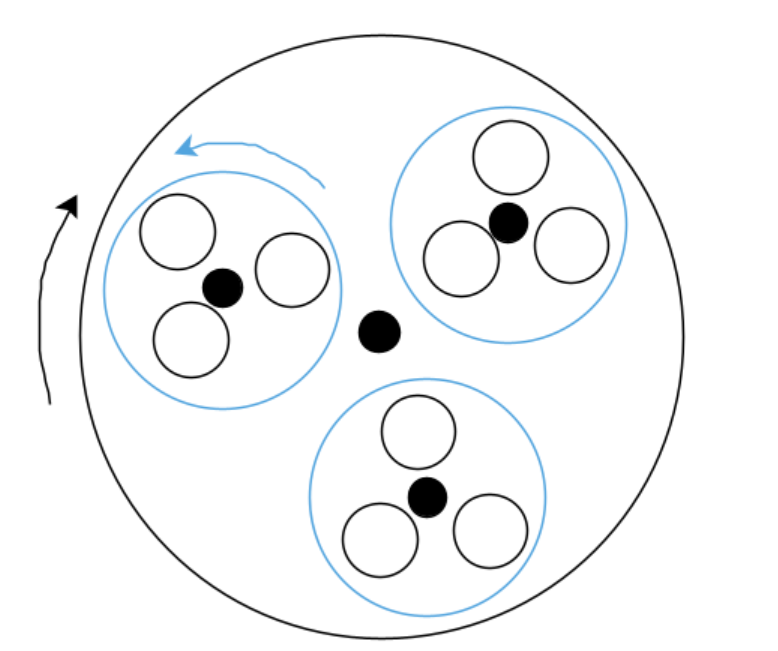}
    \caption{The model of the teacup carousel consists of a large clockwise rotating platform on top of which are smaller platforms rotating counterclockwise. The actual teacups are axially fastened to the smaller platforms and allowed to freely rotate. Passengers, or teamates, are seated on the rim of a teacup forming teacup-teamate complexes. In the model considered here, the teacups are assumed massless and hence the positions of the teamates determine the center-of-mass of the complex.}
    \label{fig:illustration}
\end{figure}

Figure~\ref{fig:illustration} illustrates teacup carousel model that we examine in this article. It consists of a large rotating platform that contains smaller platforms rotating in opposite direction. The teacups are on top the the smaller platform free to rotate independently. Together these three distinct levels of rotational motion form complex dynamics, which we will model using both analytical and numerical calculations.

The larger platform rotates at angular frequency $\Omega_0$ causing the centers of the smaller platforms follow the trajectory:
\begin{equation}
    \mathbf{r}_1(t)=r_1 \begin{pmatrix}
        \cos\Omega_0t \\ \sin\Omega_0t\\
    \end{pmatrix}.
\end{equation}
Similarly the smaller platforms rotate at angular frequency $\omega_0$, yielding the trajectory of each teacup in the smaller platform's reference frame
\begin{equation}
    \mathbf{r}_2(t)=r_2 \begin{pmatrix}
        \cos\omega_0t \\ \sin\omega_0t \\
    \end{pmatrix}.
\end{equation}

The position of the teacup as observed in inertial ground reference frame is now
\begin{equation}
    \mathbf{R}(t)=\mathbf{r}_1(t)+\mathbf{r}_2(t).
\end{equation}
It is the combination of the two rotational movements that leads to the jerky movement of the teacup carousel. The acceleration of a teacup is
\begin{equation}
    \mathbf{R}''(t)=-\Omega_0^2\mathbf{r}_1(t)-\omega_0^2\mathbf{r}_2(t).
\end{equation} 

The double rotation of the platforms provides the non-inertial reference frame for the passengers', with the position pseudovector of the passenger ${\bf r}(t)$ measured with respect to the center of the teacup. The non-inertial reference frame results in a fictitious force acting on the passenger $-m\mathbf{R}''(t)$. With the passengers sitting in the teacup, the only possible movement is rotation about the cup's center, and hence we examine the fictitious pseudotorque
\begin{equation}
    \tau=\mathbf{r}(t)\times(-m\mathbf{R}''(t)),
\end{equation}
where $m$ is the mass of the teamate.
Because the passenger is sitting in the teacup, we can assume that, at least momentarily, friction is keeping them still on their seat. This means that the pseudotorque acting on the passenger will cause the teacup to rotate. If there are multiple teamates sitting in the teacup, forming an extended teacup-teamate-complex, the total torque on the cup is the sum of the pseudotorques acting on the passengers:
\begin{equation}
    \tau_\mathrm{cup}=\sum_i\mathbf{r_i}(t)\times(-m_i\mathbf{R}''(t))=-\left[\sum_im_i\mathbf{r_i}(t)\right]\times\mathbf{R}''(t)=-M{\bf r}_\mathrm{cm}(t)\times\mathbf{R}''(t),
    \label{eq:torque}
\end{equation}
where $\mathbf{r_{cm}}$ is the position of the center of mass. The direction of this torque is such that it tries to rotate the center of mass $\mathbf{r_{cm}}$ towards the direction of the force $-M\mathbf{R}''(t)$.

\begin{figure}[h]
    \centering 
    \includegraphics[width=0.8\columnwidth]{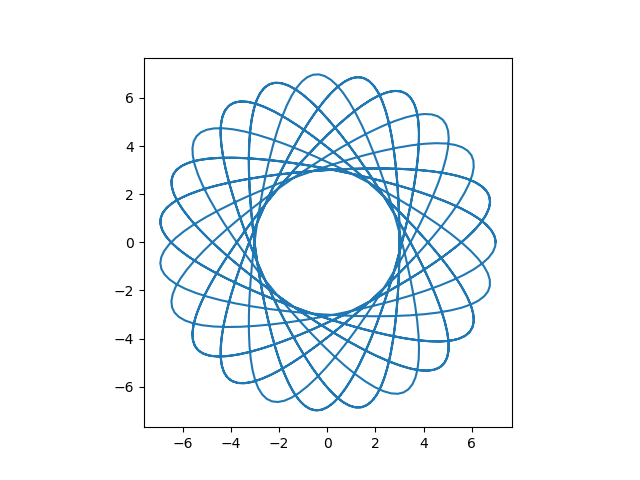}    
    \caption{Track of the center point of the teacup for $\Omega_0 = 0.2\,\mathrm{rad/s}$ and $\omega_0 = -0.24\,\mathrm{rad/s}$.}    
    \label{fig:track}
\end{figure}

Figure~\ref{fig:track} models the route of the teacup when the bigger and smaller platforms of the carousel are spinning in opposite directions. This graph does not show how the motion changes when the teacup itself is spinning. The track of the teacup resembles a flower, formed of elliptical lines.

\section{Numerical solution}

In this section we aim to find the angular acceleration. Moreover, we analyze who is attracting whom.
The previously mentioned pseudotorque in Eq.~\eqref{eq:torque} causes the teacup to undergo angular acceleration $\alpha$.
\begin{equation}
    I\alpha=\tau_\mathrm{cup}
\end{equation}
 where $I$ is the moment of inertia of the teacup-teamate-complex relative to the center of the teacup:
\begin{equation}
    I = I_0 + \sum_i m_i r_i^2 = I_0 + Mr^2,
\end{equation}
where $I_0$ is the moment of inertia of an empty teacup and $|{\bf r}_i| = r$ holds for all $i$ where $r$ is the radius of the teacup. From this we can derive the equation of motion for angular acceleration of the complex
\begin{equation}
    I\frac{d^2 \theta(t)}{dt^2} = -M {\bf r}_\mathrm{cm} (t) \times {\bf R}''(t),
\end{equation}
yielding the angular acceleration
\begin{equation}
    {\bf \alpha} = -\frac{M{\bf r}_\mathrm{cm} (t) \times {\bf R}''(t)}{I_0 + Mr^2} = -\frac{M}{I_0+Mr^2} {\bf r}_\mathrm{cm} (t) \times {\bf R}''(t).
    \label{eq:angular-acceleration}
\end{equation}

From the equation of motion, it can be trivially seen~\cite{jerkyderivation} that the final torque experienced by teamates is
\begin{equation}
{\bf \tau} = \left[-m + \frac{mM}{I_0+Mr^2}\left[{\bf r} \cdot {\bf r}_\mathrm{cm} \right] \right]{\bf r}\times{\bf R}''(t) - \frac{mM}{I_0+Mr^2} \left[{\bf r} \cdot {\bf R}''(t) \right] {\bf r}\times {\bf r}_\mathrm{cm}.
\label{eq:jerkyequation}
\end{equation}
The combined jerky movement of the double rotating platform, together with the complex dynamics of the teacup-teamate complex, provides thus additional jerks with seemingly random behaviour that is particularly attractive in the spinning teacup attraction.

Moreover, the pseudotorque acting on the teacup-teamate-complex tends to rotate its center of mass in the direction of the force. In the same manner, a pseudoforce of the same direction is acting on the passengers, leading into a force towards the center of mass, which lies in the direction of the other teamates. Therefore, on average, the teamates experience an attractive force. It is this non-inertial reference frame induced attractive interteamate interaction that was empirically observed in the teacup ride.

\begin{figure}[h!]
    \centering \includegraphics[width=0.9\columnwidth]{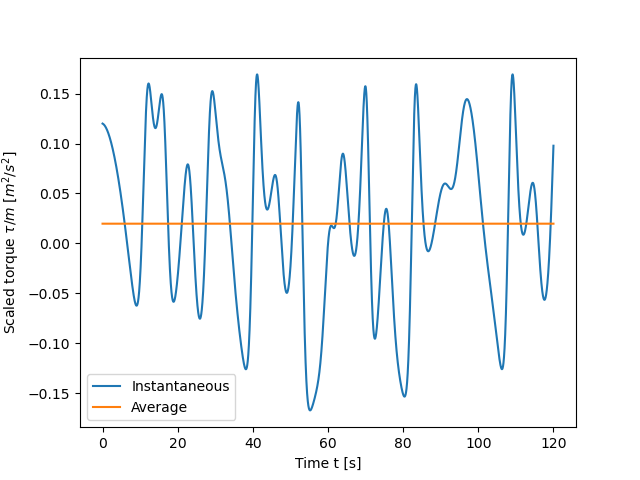}
    \caption{The instantaneous and average pseudotorque applying on the teamate for the parameters used in Fig.~\ref{fig:track} and assuming vanishing teacup moment of inertia $I_0$. Positive values of the torque corresponds to the direction towards the common center of mass and hence towards the fellow teamate. While instantaneous torque has rapid changes and sign changes, on average the torque is positive. The other teamate has identical profile but the direction is reversed.}
    \label{fig:torque}
\end{figure}

For numerical simulation of the pseudotorque in Eq.~\eqref{eq:jerkyequation} see Fig.~\ref{fig:torque}. The data shows the rapid changes in the torque, resulting in an angular jerk that is the key component that makes the ride fun and interesting. However, the average torque is positive, meaning that the time-averaged torque tends to take the teamate towards the common center-of-mass of the teacup teamate complex and hence towards the other fellow teamates in the teacup. Thus the numerical simulation verifies the empirical observation that passengers on the teacup ride experience a force that attracts them together.

\section{Conclusions}

This attractive interaction between the teamates in the teacup carousel could be theorized to be caused by a field of interaction. Like other fields of physical interaction, this field would be mediated by a mediator particle. By inspecting the properties of this mediator particle, we might discover the hidden secrets of attraction and that the science behind attraction is neither chemistry nor biology, but physics. Questions arise: does attraction travel at the speed of light and could this new perspective shed light on the long-standing debate about experiencing attraction to someone at first sight?

Physicists have generalized jerk into functions used to describe chaotic systems~\cite{sparavigna2015}. It wouldn't be far-fetched to say that people generally do not find unpredictable and chaotic behavior agreeable. Even though jerks appear to cause the previously modeled event that resembles attraction, one could conclude that this chaotic behavior typical of jerks is quite unattractive, especially in larger doses.
 
 Although this article has made advances in uncovering the connection between jerks and attraction, further study on the subject is essential. Studying jerks offers intriguing possibilities for expanding our understanding of both physical and social dynamics. Teaching more about the interplay between attraction and jerkiness could benefit university students, especially considering the absence of jerk in university-level physics textbooks.


\begin{thebibliography}{99}
\bibitem{jerkyderivation} The jerky derivation will be presented elsewhere.
\end{thebibliography}
\end{document}